\documentclass [prb,twocolumn,superscriptaddress,showkeys,preprintnumbers]{revtex4} 

\usepackage{xspace}
\usepackage {graphicx}

\begin{document}

\title{Progress in Three-Dimensional Coherent X-Ray Diffraction Imaging}

\author{S. Marchesini} \email{smarchesini@llnl.gov}
\author{H. N. Chapman}
\affiliation{University of California, Lawrence Livermore National
  Laboratory, 7000 East Ave., Livermore, CA 94550, USA}
\affiliation{Center for Biophotonics Science and Technology, UC Davis, 2700 Stockton Blvd., Ste 1400, Sacramento CA, USA}
\author{A. Barty}
\author{A. Noy}
\author{S. P. Hau-Riege}
\author{J. H. Kinney}
\affiliation{University of California, Lawrence Livermore National
  Laboratory, 7000 East Ave., Livermore, CA 94550, USA}
\author{C. Cui}
\author{M. R. Howells}
\author{R. Rosen}
\affiliation{Advanced Light Source, Lawrence Berkeley National Laboratory,
  1 Cyclotron Road, Berkeley, CA 94720, USA}
\author{J. C. H. Spence}
\author{U. Weierstall}
\affiliation{Department of Physics and Astronomy, Arizona State University,
  Tempe, AZ 85287-1504, USA}
\author{D. Shapiro}
\affiliation{Center for Biophotonics Science and Technology, UC Davis, 2700 Stockton Blvd., Ste 1400, Sacramento CA, USA}
\author{T. Beetz}
\author{C. Jacobsen}
\author{E. Lima}
\affiliation{Department of Physics and Astronomy, Stony Brook University,
  Stony Brook, NY 11794-3800, USA}
\author{A. M. Minor} 
\author{H. He}
\affiliation{National Center for Electron Microscopy, Lawrence Berkeley National Laboratory,   1 Cyclotron Rd, Berkeley, CA 94720, USA}

\begin{abstract}
The Fourier inversion of phased coherent diffraction patterns offers
images without the resolution and depth-of-focus limitations of
lens-based tomographic systems. We report on our recent experimental
images inverted using recent developments in phase retrieval
algorithms, and summarize efforts that led to these accomplishments.
These include ab-initio reconstruction of a two-dimensional test
pattern, infinite depth of focus image of a thick object, and its
high-resolution ($\sim10$ nm resolution) three-dimensional image.
Developments on the structural imaging of low density aerogel samples
are discussed.
\end{abstract}

\preprint{UCRL-PROC-215874}
\maketitle

\section{Introduction }

In the last five years or so several new ideas have combined to provide us 
with a working solution to the phase problem for non-periodic objects. This 
capability opens exciting possibilities for using coherent x-ray diffraction 
microscopy (CXDM) for 3D imaging of few-micron-sized objects at resolution 
levels previously inaccessible to x-ray microscopy. Since the first proof of 
principle demonstration of CXDM \cite{Miao:1999}, a number of groups 
have been working to bring these possibilities into reality. Recent 
estimates \cite{Howells:2004} of the dose and flux requirements of 
such measurements, indicate that attractive resolution values (about 10 nm 
for life science and 2--4 nm for material science) should be possible with 
reasonable exposure times using modern synchrotron-radiation sources. Thus 
CXDM promises a 3D resolution limited only by radiation damage, the 
collection solid angle and the number of x-rays collected. We therefore 
expect to have an advantage over lens-based tomography schemes that are 
currently limited in resolution and efficiency by the lens fabrication 
technology and, in principle, by the depth of focus effect. This capability 
provides an extremely valuable tool for understanding nanoscience, such as 
the study of minimum energy pathways for crack propagation in brittle 
solids, and characterizing the internal structure of mesoporous structures 
that are synthesized for a wide range of applications.

In this paper we review the historical developments which have led to these 
opportunities and describe some of the activities of our multi-institutional 
collaboration, working at beam line 9.0.1 at the Advanced Light Source at 
the Lawrence Berkeley National Laboratory. In particular we will describe 
here two experiments, which demonstrate spectacular 3D imaging at 10 nm 
resolution. The portion of the work devoted to life-science imaging, largely 
by the Stony Brook group, has been reported in these proceedings by Lima. 
\cite{Lima:2005}

\section{Conceptual History}

The observation by Sayre in 1952 \cite{Sayre:1952} that Bragg 
diffraction under-samples the diffracted intensity pattern was important and 
led to more specific proposals by the same author for X-ray diffractive 
imaging of non-periodic objects. \cite{Sayre:1980} These ideas, 
combined with the rapid development of computational phase retrieval in the 
wider optics community especially the ``support constraint'' 
\cite{Fienup:1978,Fienup:1980,Fienup:1982}, 
enabled the first successful use of CXDM. An important review, which
attempted to integrate the approaches of the optical and
crystallographic communities, appeared in 1990. \cite{Millane:1990}
The connection was made between the "solvent-flattening" or
"density-modification" techniques of crystallography and the compact
support requirements of the hybrid input-output (HIO) algorithm.  The
importance of fine sampling of the
\textit{intensity} of the measured diffraction pattern 
was recognised at an early stage \cite{Bates:1982} and has led to the
method being referred to as "oversampling", since the Shannon sampling
interval is half the Bragg interval in each dimension. The result of
Shannon sampling the intensity is that the diffracted \textit{phased
amplitude} is at least two-fold oversampled in each dimension, which
implies that the object obtained by transformation of an exactly-known
amplitude function will be surrounded by a zero-padded region of at
least three times the object area for 2D reconstructions, or at least
seven times the object volume for 3D reconstructions. Such zero
padding is a necessary concomitant of the use of a support constraint,
although it is sufficient to oversample at less than the factor of two
per dimension
\cite{Miao:1998}. The use of the support constraint as a means to 
phase the diffraction pattern is now very widespread in the growing CXDM 
community. 

\begin{figure}[htbp]
  \centering
  \includegraphics[height=0.3\textwidth]{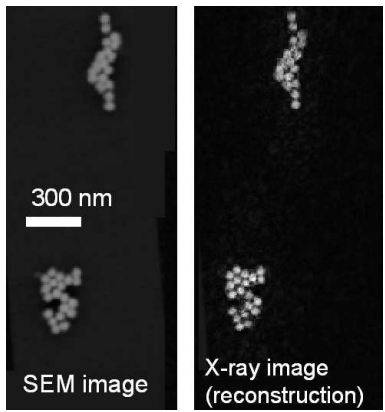}
  \caption{SEM images of gold ball clusters (left) and reconstructed soft X-ray image (right) recorded at  $\lambda=2$ nm. \cite{Marchesini:2003}}
  \label{fig:1}
\end{figure}

\section{Experiments With X-rays}

In spite of the promise shown in simulations, the above theoretical advances 
were not accompanied by immediate experimental progress in the practical 
application of phase retrieval. The first successful X-ray technique was 
developed by the Stony Brook group at the X1 undulator beam line at the 
National Synchrotron Light Source at Brookhaven. The fruit of this effort, 
reported by Miao, Charalambous, Kirz and Sayre in 1999, 
\cite{Miao:1999} was the first inversion of an experimental 
X-ray diffraction pattern to an image of a non-periodic object at 75 nm 
resolution. This success proved to be the beginning of a significant 
expansion in interest in CXDI in the US.

In the last few years CXDM activities in the US has involved four groups 
which have all made contributions to the XRM 2005 conference: Stony 
Brook/Brookhaven, University of California at Los Angeles (UCLA), University 
of Illinois / Argonne, and ourselves at University of Arizona / Livermore 
Lab / Berkeley Lab. Stony Brook / Brookhaven have constructed a 
sophisticated experimental station for tomographic imaging of life-science 
specimens at cryogenic temperatures. \cite{Beetz:2005} This apparatus 
is now installed at ALS beam line 9.0.1 \cite{Howells:2002} and 
serves all of the groups doing CXDM at the ALS. Robinson and coworkers at 
the University of Illinois have applied the principles of CXDM to hard x-ray 
experiments on microcrystalline particles, the density variations of which 
produce a diffraction pattern centered on each Bragg spot. The pattern can 
be reconstructed in 2D \cite{Robinson:2001} or scanned in 3D by very 
slight rotation of the crystal to give the equivalent of a tilt series. Such 
data have been reconstructed tomographically to produce a 3D image at 80 nm 
resolution. \cite{Williams:2003} Miao (now at UCLA) and coworkers have made 
considerable progress in pushing the CXDI method to higher resolution in 2D 
(7 nm), higher x-ray energies and to a limited form of 3D. 
\cite{Miao:2002} They have also made the first application of CXDM to 
a biological sample. \cite{Miao:2003}

Our own efforts in this area have concentrated on retrieving phase based on 
the diffraction data alone \cite{Marchesini:2003}, three dimensional 
\textit{ab-initio} reconstruction of a test 3D object made of 50 nm gold balls deposited on a 
pyramid shaped silicon nitrade membrane at 10 nm resolution, 
\cite{Chapman:2005} two dimensional images with infinite depth of 
focus, and the imaging of materials sciences samples such as Ta$_{2}$O$_{5}$ 
aerogel foams.

\section{Experimental Results}

\begin{figure}[tbp]
  \centering \includegraphics[width=0.45\textwidth]{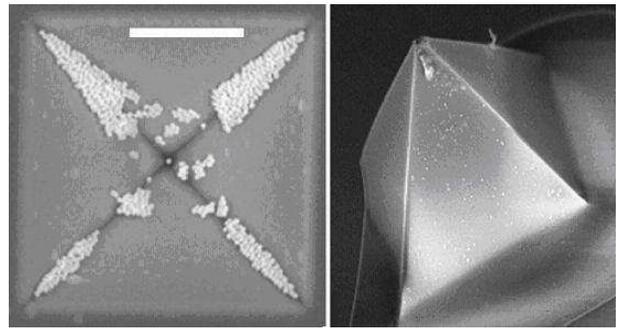}
  \caption{Scanning Electron Microscopy (SEM) of a silicon nitride
  pyramid (right) and the gold balls deposited on the hollow side of
  the membrane (left).  Scalebar is 1 $\mu$m.}  \label{fig:2}
\end{figure}
 
\begin{figure}[bp]
  \centering \includegraphics[width=0.4\textwidth]{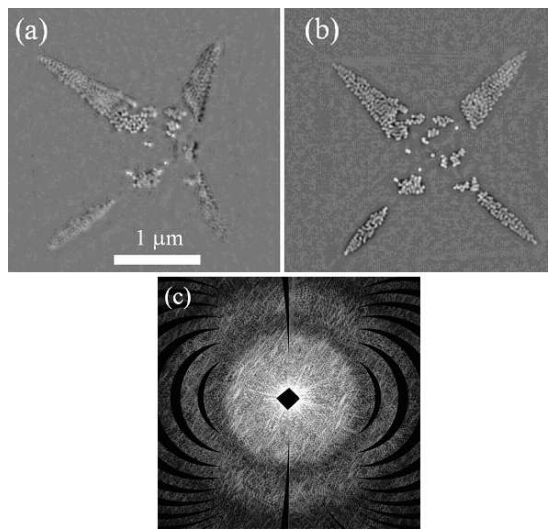}
  \caption{(a) Reconstruction from a single view diffraction pattern
  at an object orientation $\varphi = 24^\circ$.  Scalebar is 1 $\mu$m. 
(b) Infinite
  depth of focus projection images, for the object orientation $\varphi =
  0^\circ$. (c) 2D section of the three dimensional diffraction pattern: as
  the object is rotated, the recorded Ewald spheres intercept the
  plane leaving some gaps of missing data. } \label{fig:3}
\end{figure}

Our experiments reported here were carried out using the abovementioned 
 Stony Brook University diffraction apparatus, as well 
as an earlier apparatus \cite{Marchesini:2004}. In 
the Stony Brook apparatus we carried out experiments using 750 eV (1.65 nm 
wavelength) X-rays that were selected from the undulator spectrum by a 
zone-plate monochromator with a spectral resolution of $\lambda / \Delta 
\lambda $=1000. The 4-$\mu $m-diameter monochromator exit pinhole also 
selects a transversely spatially coherent patch of the beam. The
sample was located 20 mm from this pinhole. A direct-detection bare
CCD detector, with 20 $\mu $m pixel spacing, 1340$\times $1300 pixels,
was located 142 mm behind the sample. At these CCD and wavelength
settings we have an object sampling interval in $x$ and $y$ of $\Delta
x$ = 9.8 nm (in the small-angle approximation) and a field width of $w
= N\Delta x$= 11.7 $\mu $m. A beam-stop blocks the direct undiffracted
beam from impinging on the CCD.  More details are given by Beetz et
al. \cite{Beetz:2005}.  Diffraction patterns were collected with
the sample oriented at rotation angles \textit{$\phi $} of -$60^\circ$ to
$+60^\circ$, at $1^\circ$ intervals. Total exposure time was about 3 hours
per sample.

Test samples were made by placing a droplet of solution containing 
unconjugated colloidal gold balls on a silicon nitride membrane (thickness 
100 nm) and allowing it to dry. A two dimensional object was imaged without 
prior knowledge about its shape by periodically updating the support region 
based on the current object estimate (Fig. \ref{fig:1}) 
\cite{Marchesini:2003}.

A three-dimensional test sample was produced by placing a droplet of
colloidal gold solution on a three-dimensional silicon nitride
pyramid-shaped membrane. \cite{Chapman:2005} This drop quickly
evaporated and left the gold balls in a characteristic pattern where
the gold tended to fill in the edges of the pyramid. An SEM image of
the object is shown in Fig. \ref{fig:2}. The pyramid base width is 2.5
$\mu $m and the height (base to apex) is 1.8 $\mu $m. An earlier,
larger, silicon nitride pyramid object is shown on the right side of
Fig. \ref{fig:2}.

Two dimensional projection images may be recovered from the diffraction 
intensities without having to first undergo a full 3D reconstruction, and we 
found this is a useful step to quickly examine our 3D datasets. The 
diffraction intensities from a single sample orientation are recorded on the 
Ewald sphere and will have the same depth of focus as a microscope with NA 
equivalent to the solid angle intercepted by the CCD. For our experimental 
parameters, giving NA = 0.084, we have a depth of focus of 120 nm, which is 
considerably smaller than the 1.8 $\mu $m thickness of the pyramid object. A 
reconstructed image, from a single-view diffraction pattern is shown in Fig. 
\ref{fig:3}(a). Artifacts due to defocus are clearly seen in the image. By the choice 
of the parabolic term of the retreived phases \cite{Marchesini:2005} 
the plane of focus can be numerically scanned throughout the depth of the 
object.

By sectioning the three dimensional Fourier space with a sphere of
larger radius than the Ewald surface, we can increase the depth of
focus. Infinite depth-of-focus two-dimensional projection images were
obtained (Fig \ref{fig:3}b) from a plane in reciprocal space
perpendicular to the projection direction (Fig. \ref{fig:3}c)
\cite{Chapman:2005}.

\begin{figure}[htbp]
  \centering \includegraphics[width=0.4\textwidth]{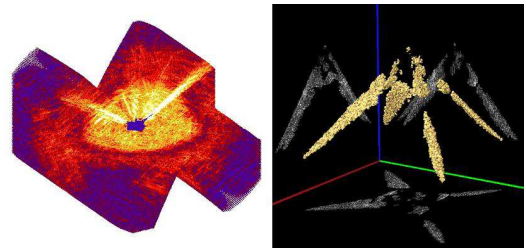}
  \caption{Three dimensional diffraction pattern (left) (with a
  quadrant removed for visualization) and reconstructed 3D images
  \cite{Chapman:2005} (right) showing the isosurface as well as the
  projection images of the sample. } \label{fig:4}
\end{figure}

\begin{figure}[htbp]
  \centering \includegraphics[width=0.4\textwidth]{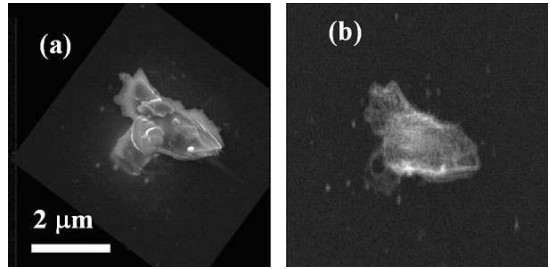}
  \caption{(a) SEM of an areogel sample with reference points used to
  help the phase retrieval, and (b) two dimensional projection of the
  reconstructed 3D image at 14 nm resolution.}  \label{fig:5}
\end{figure}

A full 3D image 
was obtained \cite{Chapman:2005} by performing phase retrieval 
\cite{Fienup:1982,Luke:2005} on the entire 3D 
diffraction dataset. The reconstructed volume image reveals the structure of 
the object in all three dimensions and can be visualized in many ways 
including isosurface renderings, projections through the data (Fig. 
\ref{fig:4}), or slices (tomographs) of the data. 

We have applied 3D diffraction imaging to determining the 3D structure
of low density aerogel foam samples. These metal oxide foams are low
density (100 mg/cc) and have an internal skeleton structure composed
of Ta$_{2}$O$_{5}$. Our full 3D reconstructions were obtained with no
a priori information about the sample, including no measurement of the
missing low spatial frequency data in the beamstop region. The
reconstructed image, shown in Fig. \ref{fig:5}, reveals not only the
particle shape, but also internal foam structure such as the strut
geometry, which can be used to calculate the foam mechanical
properties. Further details will be given in another paper.

\section{Phase Retrieval and Image Analysis}

We overcame two key computational challenges in implementing
high-resolution 3D phase retrieval, \cite{Barty:1} specifically
performing the numerous 1k$^{3}$ FFTs required for phase retrieval in
a reasonable time and managing the memory requirements of large 3D
data sets. Memory and calculation requirements are significant and
suggest a cluster-based solution. The FFTs (dist{\_}fft) have been
optimised for the G5 vector processor architecture by the Apple
Advanced Computation Group and uses standard MPI interfaces to perform
distributed giga-element or larger FFTs. Reconstruction code is
written in C, is fully parallelised, and uses distributed memory and
MPI interfaces to share the workload across all CPUs in the
system. This includes application of real and Fourier space
constraints and dynamic support refinement using the Shrinkwap
algorithm.

Using 16 G5 Macintosh computers with Infiniband interconnects we achieve an 
FFT speed of 7.9 sec/FFT on a 1024$^{3}$ voxel data set, giving a full 
reconstruction time of 14 hours (based on 2000 iterations, 2 FFTs per 
iteration plus other floating point operations needed for the 
reconstruction). Timings for a 512$^{3}$ data cube are 850 msec/FFT, 
enabling us to perform a full reconstruction in 1.5 hrs.

\begin{figure}[htbp]
  \centering \includegraphics[width=0.4\textwidth]{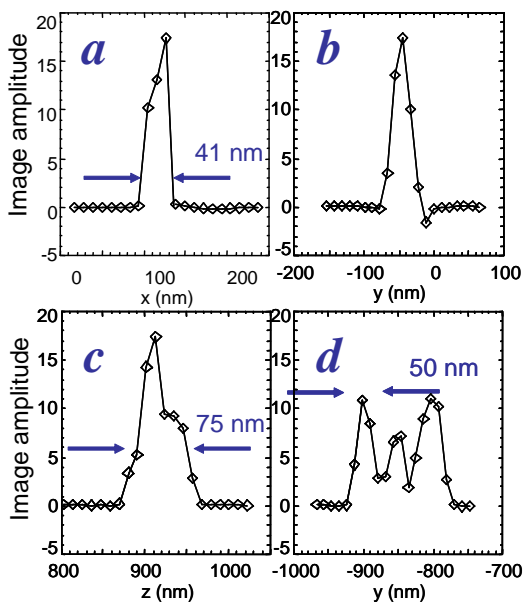}
  \caption{Line-outs of the real part of the reconstructed complex
  amplitude 3D image, through the isolated single ball at the pyramid
  apex. Dashed lines show a simulated 3D coherent image with a cube
  OTF of 10 nm resolution and with a 60$^\circ$ missing sector
  \cite{Chapman:2005}.  The lineout through three lines (d) demonstrates
  that the resolution is sufficient to clearly distinguish between
  different 50 nm gold spheres.}  \label{fig:6}
\end{figure}

Although we cannot exactly quantify the resolution of the image, which would 
require knowing the object's 3D structure, our analysis shows we can 
consistently retrieve phases out to the maximum spatial frequency recorded 
\cite{Chapman:2005} (further improvements in. 
\cite{Marchesini:2005} A line-out through the reconstructed image can 
easily resolve 50 nm spheres that are touching each other 
(see Fig. \ref{fig:6}). From such image line-outs, and 
comparisons of reconstructed X-ray images with the SEM image, we have 
confidence that our achieved image resolution is close to 10 nm. Further 
analysis of the consistency of the retrieved phases, and the agreement of 
the Fourier amplitudes of the reconstructed image with the data, back up 
this assertion\cite{Chapman:2005,Marchesini:2005}.

\section{Holographic-Enhanced Phase Retrieval }

\begin{figure}[htbp]
  \centering \includegraphics[width=0.35\textwidth]{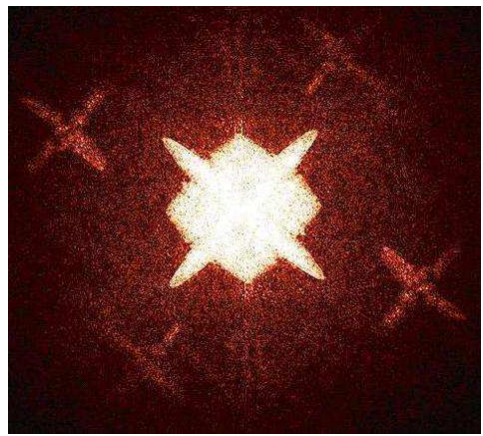}
  \caption{Fourier transform of the diffraction pattern of the object
  illuminated with a large beam: the central part of the picture
  shows the autocorrelation of the pyramid, but some reference points
  produce off-centered holograms. } \label{fig:7}
\end{figure}

It was noted that the autocorrelation functions in some of this work
also included faithful, although low-resolution, holographic images of
some of the clusters, due to the occurrence of a single isolated ball
near the object. In analogy with the ``heavy atom'' method of
crystallography, by placing a reference point object near the sample
we can obtain a one-step, estimate of the support function. Although
the holographic image is noisier than the recovered image
(Fig. \ref{fig:7}), it provides a useful starting point to the
algorithm.

\begin{figure}[htbp]
  \centering \includegraphics[width=0.45\textwidth]{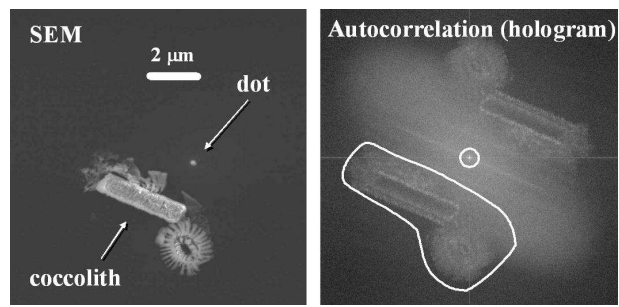}
  \caption{SEM image of a coccolith shell deposited on a silicon
  nitride membrane (left). A reference point deposited with a focused
  ion beam produces a holographic image (right).}  \label{fig:8}
\end{figure}

Inspired by this holographic method to help the phase retrieval step (see 
also \cite{Marchesini:2005}, we developed a methodology to deposit 
controlled reference points near the object by metallorganic deposition 
using a focused ion beam. Our initial trials are illustrated in Fig. 
\ref{fig:8}.

\section{Conclusions }

We have demonstrated \textit{ab-initio} two dimensional images with infinite depth of focus, 
and three dimensional images of test objects at a resolution of 10 nm. 
Preliminary images of areogel foams were presented. These images of 
complicated and unknown objects, along with the rigorous analysis of known 
test objects, show the robustness of our \textit{ab initio} phase retrieval technique. In the 
case of the aerogel particle, the reconstruction was performed ``blind'' 
without the operator (A. Barty) of the reconstruction software aware of the 
SEM image, or the size or shape of the object. 

While the recent experimental progress to date has been rapid and extremely 
encouraging, we are looking forward to further improvements in the 
technique, including faster acquisition times (with an improved beamline) 
that will allow us to achieve even higher image resolution. Given the 
scaling of required dose to the inverse fourth power of resolution 
\cite{Howells:2004}, and estimates of coherent flux improvements 
achievable with an optimized beamline and undulator source, we estimate that 
we should be able to achieve resolutions of 2--4 nm on material science 
samples. Possible \cite{Howells:2004} applications for the technique 
include characterizing the pore structure of vesicular basalt, the formation 
of voids in metals, and many other investigations of the nanoworld. The 
techniques that we have developed will also be applied to exciting new 
prospects for imaging of large macromolecules and assemblies at near 
atomic-resolution imaging, which will be achieved using X-ray free-electron 
lasers \cite{Neutze:2000} and aligned molecule 
diffraction\cite{Spence:2004}.

\acknowledgments

Coccolith samples were provided by J. Young from the Natural History
Museum, London. This work was performed under the auspices of the
U.S. Department of Energy by University of California, Lawrence
Livermore National Laboratory under Contract W-7405-Eng-48. This work
has been supported by funding from the National Science
Foundation. The Center for Biophotonics, an NSF Science and Technology
Center, is managed by the University of California, Davis, under
Cooperative Agreement No. PHY 0120999. The work of the Lawrence
Berkeley National Laboratory participants and the operation of the ALS
facility was supported by the Director, Office of Energy Research,
Office of Basic Energy Sciences, Materials Sciences Division of the
U. S. Department of Energy, under Contract No. DE-AC03-76SF00098.

\end{document}